\documentclass{elsarticle}
\usepackage{amsmath}
\makeatletter
\@addtoreset{equation}{section}
\makeatother

\usepackage{etoolbox}
\makeatletter
\def\@mkboth#1#2{}
\newlength\appendixwidth
\preto\appendix{\addtocontents{toc}{\protect\patchl@section}}
\newcommand{\patchl@section}{%
  \settowidth{\appendixwidth}{\textbf{Appendix }}%
  \addtolength{\appendixwidth}{1.5em}%
  \patchcmd{\l@section}{1.5em}{\appendixwidth}{}{\ddt}%
}
\makeatother

\begin{document}
	\begin{frontmatter}
		
		\title{Non-Standard Primordial Clocks from Dynamical Mass in Alternative to Inflation Scenarios}
		\author[1,2]{Yi Wang}
		\ead{phyw@ust.hk}
		\author[1]{Zun Wang}
		\ead{zwangdq@connect.ust.hk}
		\author[1,2]{Yuhang Zhu}
		\ead{yzhucc@connect.ust.hk}
		
		\address[1]{Department of Physics, The Hong Kong University of Science and Technology, Clear Water Bay, Kowloon, Hong Kong, P.R.China}
		\address[2]{Jockey Club Institute for Advanced Study, The Hong Kong University of Science and Technology, Clear Water Bay, Kowloon, Hong Kong, P.R.China}

		\begin{abstract}
			In the primordial universe, oscillations of heavy fields can be considered as standard clocks to measure the expansion or contraction history of the universe. Those standard clocks provide a model-independent way of distinguishing inflation and alternative scenarios. However, the mass of the heavy fields may not be a constant mass, but rather mass dynamically generated by non-minimal coupling to the Ricci scalar, or self-interactions. In the case of dynamically generated mass, the mass of the heavy field is generically of order Hubble, and thus is time-dependent in alternative to inflation scenarios. We show that such dynamically generated mass terms can be considered as non-standard primordial clocks for alternative to inflation, providing similar oscillatory frequencies as standard clocks of inflation. Additional information on scale dependence can distinguish such non-standard clocks from standard clocks.			
		\end{abstract}
	\end{frontmatter}
	\newpage

	\tableofcontents

	\section{Introduction}
	
	For the past decades, inflation has become the most popular scenario to describe the very early universe. It can solve the horizon problem and flatness puzzles in the hot big bang~\cite{ Guth:1980zm,Linde:1981mu,Albrecht:1982wi,Starobinsky:1980te}. Inflation predicts the primordial density fluctuations, which shows excellent agreement with experimental data from Large Scale Structure (LSS) and Cosmic Microwave Background (CMB).\\
	
	However, many alternative to inflation scenarios were also proposed in the literature. For example, slow contraction scenarios such as the Ekpyrosis~\cite{Khoury:2001wf}, fast contraction scenarios such as matter bounce~\cite{Wands:1998yp,Finelli:2001sr}, and slow expansion scenarios such as~\cite{Piao:2003ty}, string gas cosmology~\cite{Brandenberger:1988aj,Nayeri:2005ck}, or Galilean genesis~\cite{Creminelli:2010ba,Wang:2012bq}. \\
	
	For both inflation and alternative to inflation scenarios, a large number of models can be built, with a wide spectrum of predictions. To figure out the expansion (or contraction) history of the universe, and distinguish inflation from alternatives, it may be more efficient to start not from the model details, but rather model-independent features purely from the time-dependence of the scale factor, which is the defining features of inflation or an alternative scenario. One example is the primordial gravitational waves \cite{Starobinsky:1979ty} which are the tensor part of the primordial fluctuation, and it may be observed in the CMB experiment from B-mode polarization. Another possible way is to use massive fields that existed in the primordial universe~\cite{Chen:2009we,Chen:2009zp,Baumann:2011nk,Arkani-Hamed:2015bza}. The classical \cite{Chen:2011zf, Chen:2011tu} or quantum \cite{Chen:2015lza, Chen:2016cbe, Chen:2016qce} oscillations of massive fields work as primordial standard clocks which can directly record the time evolution of the scale factor $a(t)$. 
	
	In primordial standard clocks, the mass parameter is regarded as time-independent. However, non-standard clocks with time dependent mass are also discussed \cite{Chen:2011zf, Huang:2016quc, Domenech:2018bnf}. For the case of inflation, though it is natural to consider time-independent mass due to the approximate scale invariance of inflation, exceptions arise from mass terms arising from interactions~\cite{Domenech:2018bnf}, or interactions without a mass term~\cite{Huang:2016quc}. For alternative scenarios, in addition to a fixed mass parameter, a mass of order Hubble parameter may be more naturally generated in a dynamical way. The Hubble parameter is strongly time-dependent in alternative to inflation, and thus provides significant time dependence to the mass of the field.\\
	 
	In this work, we study those non-standard clocks with time-dependent mass in alternative scenarios. Generally, the time-dependent mass can be generated from two different mechanisms. Firstly, the massive fields have the non-minimal coupling with Ricci curvature $R$ such as $\xi R\sigma^2$. In the inflation case, the background is approximately de-sitter geometry and thus Ricci scalar contributes to the effective mass time-independently (proportional to $\xi H ^2$). However, in alternative to inflation scenarios, this term can generate a time-dependent mass. Another mechanism is from quantum loop correction. As we know in de-sitter case, with a self-interaction $\lambda \sigma^4$, due to the strange IR behaviors it can generate an effective mass $m_{\mathrm{eff}}^2\propto  H^2$ for scalar fields~\cite{Starobinsky:1994bd,Finelli:2008zg,Burgess:2009bs,Burgess:2010dd,Rajaraman:2010xd,Beneke:2012kn,Marolf:2010zp,Chen:2016nrs,Chen:2016uwp,Chen:2016hrz}. By introducing this simple self-interaction $\lambda \sigma^4$ into alternation to inflation scenarios, it may also generate a time-dependent effective mass. Following the standard procedure to calculate correlation functions, we show that with a time-dependent mass, the clock signal is different from previous works.
	\\

	The paper is organized as follows: We briefly review the concept of quantum standard clocks in section~\ref{sec2}. Then in section~\ref{sec3} and~\ref{sec4} we discuss the quantum non-standard clock signal with time-dependent mass generated from Ricci scalar and self-interaction term respectively. We summarize in section~\ref{con}. 
	
	\section{Quantum standard clocks}\label{sec2}
	\subsection{Quantum fluctuations of massive fields}
	Cosmological scale factors are expressed as power law functions with different index $p$ to describe possible conditions,
	\begin{align}
	a(t) = a_0     \left(\frac{t}{t_0}\right)^p~, 
	\end{align}
	where $a_0$ and $t_0$ are constants. The Hubble parameter is thus $H = \frac{\dot{a}}{a} = \frac{p}{t}$.
	In arbitrary FRW time-dependent background, the equation of motion of massive scalar field is
	\begin{align}
	\delta \ddot \sigma + 3H \delta \dot \sigma - \frac{1}{a^2} \partial_i^2 \delta\sigma + m^2 \delta\sigma = 0~, 
	\end{align}
	or expressed by the conformal time $\tau$, which is related to t by $a \tau = \frac{t}{(1-p)}$,
	\begin{align}\label{masseom}
	\delta\sigma'' + 2Ha \delta\sigma' - \partial_i^2 \delta\sigma + m^2 a^2 \delta\sigma = 0~.
	\end{align}
	We are interested in the massive fields in the classical regime where the long wavelength mode satisfies $\frac{k}{a} \ll m$. And the equation of motion can be approximated as
	\begin{align}
	\ddot v_k + 3H \dot v_k + m^2 v_k = 0~. 
	\end{align}
	This classical-like behavior is necessary for the quantum standard clock for contraction scenarios.
	\subsection{Quantum standard clocks}
	As the standard clock, the massive field oscillates to record the evolution of scenarios. The curvature mode resonates with the massive clock field when their wavenumbers match with each other. In an arbitrary time-dependent background, the mass of the massive scalar field is larger than scale of horizon, which causes  oscillation as $ \propto e^{\pm i m t}$ with approximated constant frequency.\\

	The quantum fluctuation of the massive field occurs as the standard clock in the time-dependent background. It can generate the clock signal with curvature modes $u_k$ with the oscillation $e^{ -iK\tau}$. In order to clarify the properties of it, we consider the two coupling terms between the perturbation of curvature field $\zeta$ and clock field $\delta \sigma$. The bilinear coupling is used as $L_2 \propto c'_2 a^3 \delta\sigma \dot{\zeta} $ and the cubic coupling is $L_3 \propto c'_3 a^3 \delta\sigma \dot{\zeta}^2 $, where $c_2'$ and $c_3'$ represent coupling constants.\\
	
	By using the in-in formalism \cite{Weinberg:2005vy, Chen:2010xka, Wang:2013zva}, the three-point function of the curvature perturbation can be written as \cite{Chen:2015lza}
	$$ \langle \zeta^3 \rangle ' \ \supset \  \int_{t_0}^{t} d \tilde{t_1} \int_{t_0}^{t} dt_1\langle0|H_I(t_1)\zeta_I^3 H_I(\tilde{t_1})|0\rangle'  $$
	$$- 2\mathrm{Re}\left(\int_{t_0}^{t} dt_1 \int_{t_0}^{t_1} dt_2 \langle 0|\zeta_I^3 H_I(t_1) H_I(t_2)|0 \rangle'\right)$$
	\begin{align}
	= 2 u_{k_3}^{*} u_{k_1} u_{k_2} |_{\tau = 0} \left(\int_{-\infty}^{0} d\tau c_2' a^3 v_{k_3} 
	u_{k_3}'\right)\left(\int_{-\infty}^{0} d\tau c_3' a^2 v_{k_3}^{*} u_{k_1}^{'*} u_{k_2}^{'*}\right)
	\end{align}
	$$ -2 u_{k_3} u_{k_1} u_{k_2}|_{\tau = 0}$$
	\begin{align}
	\times\left(\int_{-\infty}^{0} d\tau_1 c_2' a^3 v_{k_3} u_{k_3}^{'*} \int_{-\infty}^{\tau_1} d\tau_2 c_3' a^2 v_{k_3}^{*} u_{k_1}^{'*} u_{k_2}^{'*}\right.
	\end{align}
	\begin{align} \left. +\int_{-\infty}^{0} d\tau_1 c_3' a^2 v_{k_3} u_{k_1}^{'*} u_{k_2}^{'*} \int_{-\infty}^{\tau_1} d\tau_2 c_2' a^3 v_{k_3}^{*} u_{k_3}^{'*}\right) 
	\end{align}
	$$+ \mathrm{c.c.} + 2 \ \mathrm{perms}~.$$
	Here, $u_k$ and $v_k$ denote the mode functions of $\zeta$ and $\delta\sigma$. The prime refers to the expression without the common factor $(2\pi)^3 \delta{(\Sigma_i \mathbf{p_i})}$. The term ``$2 \ \mathrm{perms}$'' represents the permutation $k_1 \leftrightarrow k_3$ and $k_2 \leftrightarrow k_3$.

	\subsection{The alternative scenarios $(0 < p < 1)$}
	The relations of scale factor can be deduced from (2.1)
	\begin{align}
	a = a_0 \left(\frac{\tau}{\tau_0}\right)^{\frac{p}{1-p}}, \ \ \tau= \frac{1}{1-p}\frac{t_0^p}{a_0} t^{1-p}~. 
	\end{align}
	When two modes resonant, massive fields satisfy the classical regime conditions and the curvature mode is still sub-horizon. For the generation of the clock signal, we need two short curvature modes at the sub-horizon region. We choose $\tau_0 = -1$ for simplified calculation. By choosing Bunch-Davies (BD) vacuum under the squeezed limit $k_3/k_{1,2} \rightarrow 0$, the mode function of curvature mode can be approximately as
	\begin{align}
	u_k  \propto \frac{1}{(-\tau)^{\frac{p}{1-p}}} e^{-i k \tau}.
	\end{align}
	
	For the case of constant mass parameter $m$, the mode function can be represent as 
	\begin{align}
	v_k \propto \frac{1}{(-\tau)^{\frac{3p}{2(1-p)}}}\left[c_{+}e^{i (1-p)a_0 m (-\tau)^{\frac{1}{1-p}}} + c_{-}e^{-i (1-p)a_0 m (-\tau)^{\frac{1}{1-p}}}\right]~.
	\end{align}
	In this equation, the mass of the field is assumed time-independent. However, when the background has no scale invariance, the mass of the field could naturally be time-dependent. The generalization $m=m(t)$ can change the predictions of the quantum clocks into non-standard ones.
	
	\section{Time-dependent effective mass from coupling to Ricci curvature} \label{sec3}

	\subsection{The effective mass}
	The non-minimal coupling between the Ricci curvature $R$ and the scalar field $\sigma$ can be considered as a time-dependent contribution to the effective mass. The total effective mass $\mu$ is
	\begin{align}
	\mu^2 = m^2 + \xi R~,
	\end{align}
	where $\xi$ is the coupling constant of the non-minimal coupling, and the original mass $m$ is a constant. The total effective mass should be a combination of these two contributions.
	The Ricci scalar of the FRW metric is
	\begin{align}
	R 
	=  6 \left[\frac{\ddot{a}}{a} + (\frac{\dot{a}}{a})^2\right]~.
	\end{align}
	We consider the scale factor $a$ as a time-dependent monomial function $a(t) = a_0 (\frac{t}{t_0})^p$ so 
	\begin{align}
	R = 6 \frac{p (2p-1)}{t^2}~,
	\end{align}
	\begin{align}
	\mu^2 = m^2 + 6\xi \frac{p(2p-1)}{t^2}~.
	\end{align}
	For alternatives scenarios with time-dependent mass, we consider the differential equation of scalar perturbation for the classical regime. We will discuss the general form in \ref{eom}. For simplification, we here focus on the parameters such that the bare mass $m$ is small, though the transition between a constant mass and a dynamical mass may be interesting, similar to the discussions in \cite{Chen:2018cgg}. Neglecting $m$, and using the relation $\tau= \frac{1}{1-p}\frac{t_0^p}{a_0} t^{1-p}$, we rewrite the time variable of the differential equation into conformal time,
	\begin{align}
	\frac{d^2 v_k}{d\tau^2}+\frac{2p}{\left(1-p\right)\tau}\frac{dv_k}{d\tau}+\left[6\xi\frac{p\left(2p-1\right)}{\left(1-p\right)^2\tau^2}\right]v_k= 0~.
	\end{align}
	The general solution is
	\begin{align}
	v_k(\tau) \rightarrow C_1\left(\frac{\tau}{\tau_{k}}\right)^{\frac{\sqrt{\frac{3}{2}}\sqrt{p\left(2p-1\right)}\sqrt \xi\left(-A+B\right)}{-1+p}} + C_2 \left(\frac{\tau}{\tau_{k}}\right)^{\frac{\sqrt{\frac{3}{2}}\sqrt{p\left(2p-1\right)}\sqrt \xi\left(A+B\right)}{-1+p}}~,
	\end{align}
	where $A$ and $B$ are
	\begin{align}
	A=\frac{1}{6}\sqrt{-144-\frac{36}{\left(2p-1\right)\xi}+\frac{6}{p\left(2p-1\right)\xi}+\frac{54p}{\xi\left(2p-1\right)}}~,
	\end{align}
	\begin{align}
	B = \frac{-\sqrt6\sqrt{\left(2p-1\right)p}+3\sqrt6p\sqrt{\left(2p-1\right)p}}{6p\left(2p-1\right)\sqrt \xi}~.
	\end{align}
	Here, $t_k$ refers to the time at $a(t_k) = k/m$ for one boundary of the resonance.
	\subsection{Appropriate range of p}
	An oscillating massive field is required for generating the clock signal, which restricts the range of $p$ in (3.7). In this case, not all alternative scenarios could generate the clock signal. With $\xi > 0$, the power index 
	\begin{align}
	\frac{ \sqrt{\frac{3}{2}}\sqrt{p\left(2p-1\right)}\sqrt{\xi}\left(\pm A+B\right)}{p-1}
	\end{align}
	must contain an imaginary part to make sure the oscillation. \\
	From (3.9), the index $B(p)$ cannot make any contributions to the imaginary part in any scenarios since 
	\begin{align}
	\mathrm{Im}\left[\sqrt{p(2p-1)} B \right]=\mathrm{Im}\left[\frac{\sqrt6\left(2p-1\right)p+3\sqrt6p\left(2p-1\right)p}{6p\left(2p-1\right)\sqrt \xi}\right] = 0~. 
	\end{align}
	The index $A$ will only remain the imaginary part when it satisfies
	\begin{align}
	-144-\frac{36}{\left(2p-1\right)\xi}+\frac{6}{p\left(2p-1\right)\xi}+\frac{54p}{\xi\left(2p-1\right)} < 0~.
	\end{align}
	We should make sure the existence of an imaginary part in $\sqrt{p(2p-1)A}$ to generate the oscillatory field. The allowed interval of $p$ depends on the value of coupling coefficient $\xi$. For $\frac{1}{6} < \xi < \frac{3}{16}$, $p$ satisfies
	\begin{align}
	\frac{4 \xi -1}{16 \xi -3}-2 \sqrt{\frac{2}{3}} \sqrt{\frac{6 \xi ^2-\xi }{(16 \xi -3)^2}}< p<2 \sqrt{\frac{2}{3}} \sqrt{\frac{6 \xi ^2-\xi }{(16 \xi -3)^2}}+\frac{4 \xi -1}{16 \xi -3}~.
	\end{align}
    For the case $\xi = \frac{3}{16}$, $p$ should be larger than 2/3. And for the cases $\xi > \frac{3}{16}$, 
    \begin{align}
    p < \frac{4 \xi -1}{16 \xi -3}-2 \sqrt{\frac{2}{3}} \sqrt{\frac{6 \xi ^2-\xi }{(16 \xi -3)^2}} \ \ \  \mathrm{or}\ \ \ p>2 \sqrt{\frac{2}{3}} \sqrt{\frac{6 \xi ^2-\xi }{(16 \xi -3)^2}}+\frac{4 \xi -1}{16 \xi -3}~.
    \end{align}
	Notice that the cases for $\xi \leq \frac{1}{6}$ is not allowed. When $\xi = \frac{1}{6}$, it leads to the conformally invariance of the Klein-Gordon equation containing the effective mass with $m = 0 $, which has no oscillatory clock signals. This is expected since a conformally coupled field does not ``know'' the FRW expansion of the universe (which is conformally Minkowski).\\
	
\begin{figure}[hptb]
			\centering 
		\includegraphics[width=0.8\textwidth]{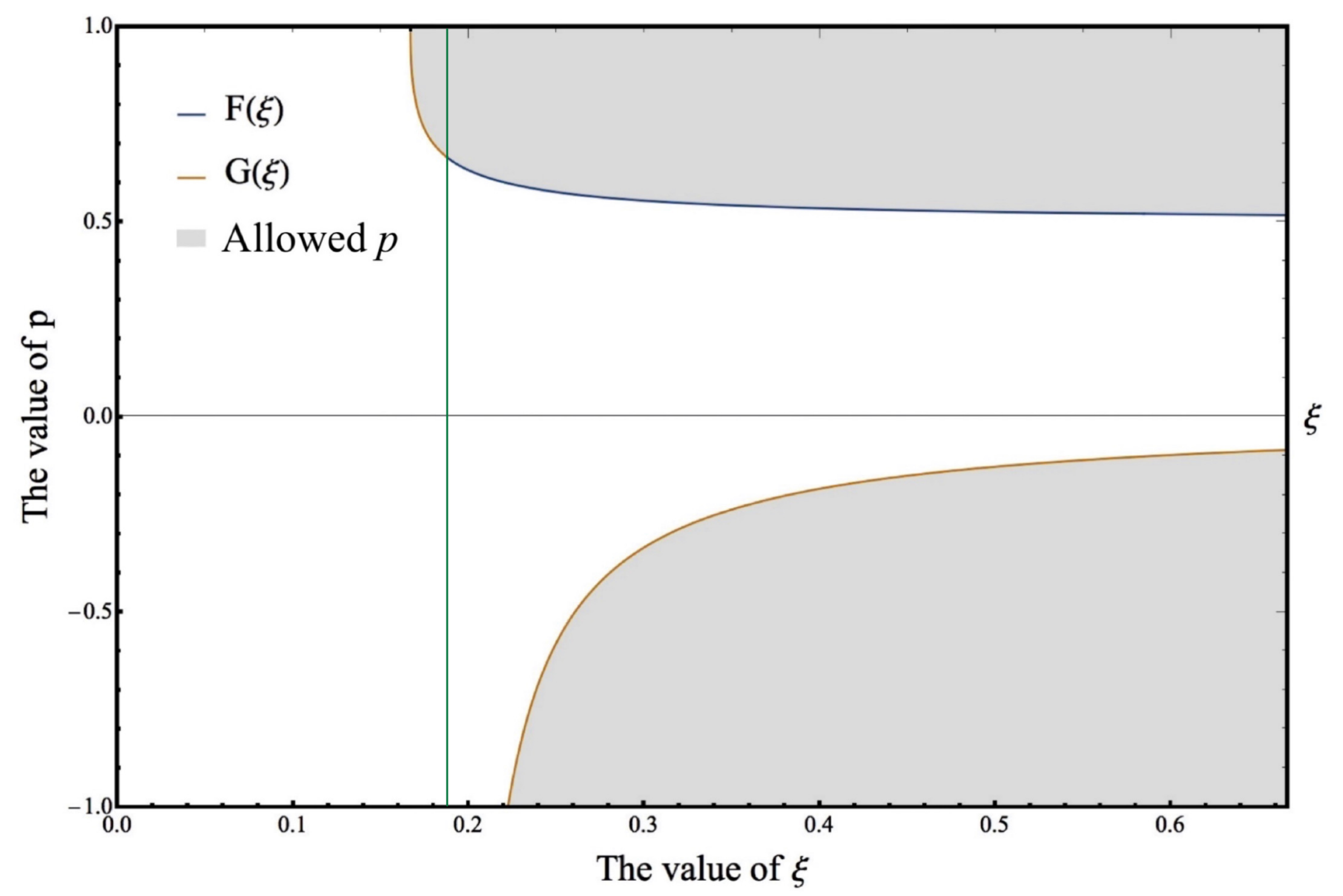}
		\caption{Choices of $p$ and $\xi$. In this figure, the gray shadow parts corresponds to the choices of $p$ and $\xi$ where there is oscillatory clock signal. The dark green line represents for the discussion boundary  $\xi = 3/16$. \label{fig1} }
	\end{figure}

	We show the allowed ranges of $p$ related to different choices of $\xi$ in Figure~\ref{fig1}. Here we only consider the scenarios with $-1<p<1$. In this figure, we redefine the R.H.S. of (3.12) as the function $F(\xi)$ and the L.H.S as $G(\xi)$, which is related to the dark yellow and blue lines in the Figure~\ref{fig1}. For $p>0$, We can figure out that the allowed $p$ should be always larger than 1/2, which is the limit for the $\xi \rightarrow \infty$ case. Thus, slow contraction scenarios will not generate such non-standard clock signals from non-minimal couplings. But the slow-expansion scenarios are allowed for some ranges of $\xi$.

	\subsection{The non-time-ordered integral}
	According to Eq.(2.9) and (3.6), the mode functions are
	\begin{align}
	v_k(\tau) \rightarrow \left( \frac{\tau}{\tau_{k}} \right)^{\frac{1-3p}{2-2p}}\left[e^{i S \mathrm{log}(-\tau)} C_1 + e^{-i S \mathrm{log} (-\tau)} C_2\right]~,
	\end{align}
	\begin{align}
	u_k(\tau) \rightarrow \frac{1}{\sqrt{k} (-\tau)^{\frac{p}{p-1}}} e^{-i k \tau}~,
	\end{align}
	where
	\begin{align}
	S = \frac{\sqrt{(48\xi - 9) p^2 + 6(1-4 \xi)p -1}}{2(p-1)}~.
	\end{align}
	For non-time-ordered integral Eq.(2.5),  the resonance among these fields can only occur in the second term inside the second parenthesis. To describe the property of this term qualitatively, we use the method of saddle point approximation which we left more details in Appendix B. For the curvature field mode $u_k$, 
	\begin{align}
	u_k \propto \frac{1}{\sqrt{k} (-\tau)^{\frac{p}{1-p}}} e^{- i k \tau} \ \ \Rightarrow \ \ u_k' \propto \frac{1}{\sqrt{k}}(-\tau)^{-\frac{p}{1-p}} [-\frac{p}{(1-p)\tau} - ik] e^{- i k \tau}~.
	\end{align}
    The most important resonance term is
	\begin{align}
	\int_{-\infty}^0 d\tau c_3' a^2 v_{k_3}^{*} u_{k_1}^{'*} u_{k_2}^{'*}~.
	\end{align}
	We notice that only the second term of Eq.(3.15) with coefficient $C_2$ can form the resonance. So the above integral should be proportional to
	\begin{align}
	\sim C_2\int_{-\infty}^0 d\tau \left(\frac{\tau}{\tau_{k_3}}\right)^{\frac{1-3p}{2-2p}}\frac{\left[-\frac{p}{(1-p)\tau}+i k_1\right]^2}{k_1} e^{i[S \mathrm{log}(-\tau)+ 2 k_1 \tau]} ~,
	\end{align}
	where the parameter $\tau_{k_3}$ is assumed to satisfy
	\begin{align}
	a(\tau_{k_3}) = \frac{k}{\mu_{\mathrm{eff}}} \ \ \Rightarrow \ \ \tau_{k_3}  = -\frac{\sqrt{6 \xi p (2p-1)}}{1-p}  \frac{1}{k_3}~.
	\end{align}
	Under the squeezed limit $k_3/k_{1,2} \rightarrow 0$, resonance happens when the conformal time $\tau_{*}$ satisfies
	\begin{align}
	\frac{d}{d \tau} (S\  \mathrm{log}(-\tau) + 2k_1 \tau) = 0~\bigg|_{\tau=\tau_{*}} \ \ \Rightarrow \ \ \tau_{*} = - \frac{S}{2} \frac{1}{k_1}~.
	\end{align}
	According to the method of saddle point approximation, the integral describing the resonance should be proportional to
	\begin{align}
	\sim C_2 \Omega\ 
	\mathrm{exp}\left\{ i S \  \mathrm{log}\left[\omega\left(\frac{2k_1}{k_3}\right)^{-1}  \right] - i S \pm  i \frac{\pi}{4}\right\}  \left(\frac{2k_1}{k_3}\right)^{\frac{3p-1}{2-2p}}~.
	\end{align}
	The parameters $\Omega$ and $\omega$ denotes
	\begin{align}
	\Omega(p, \xi) = \omega^{\frac{3p-1}{2-2p}}\sqrt{\frac{\pm \pi S}{2}}\left[i + \frac{2 p }{(1-p)S}\right]^2~, 
	\end{align}
	\begin{align}
	\omega(p , \xi) = \frac{\sqrt{6 \xi p (2p-1)}}{(1-p)S}~. 
	\end{align}
	We collect the $k_1$-independent parameters into $f(p, \xi,k_3)$ and $\phi(p, \xi, k_3)$, which depends on the details of alternative to inflation models. In general, the coefficient $C_1$ and $C_2$ may depend on $k_3$. This introduces parts of $k_3$ dependence in the prefactor $f(k_3)$ and the phase $\phi(k_3)$. We also assume $u_{k_i}\big|_{\tau=0}$ to be proportional to $1/\sqrt{k^3_i}$ .  So the non-time-ordered integral is
	\begin{align}
	\propto f(p, \xi,k_3 )\left( \frac{2k_1}{k_3}\right)^{\frac{3p-1}{2-2p}} \mathrm{exp} \left\{ i S \ \mathrm{log}\left[ \left(\frac{2k_1}{k_3}\right)^{-1}\right] + i \phi(p , \xi, k_3) \right\}~.  
	\end{align}
	Under the squeezed limit $k_3/k_{1,2}\rightarrow 0$, the clock signal is considered as the oscillatory shape function in the three function in the previous work. It is defined as
	\begin{align}
	\langle \zeta^3 \rangle \equiv S(k_1, k_2, k_3) \frac{1}{(k_1 k_2 k_3)^2} (2\pi)^7 \tilde{P}_{\zeta}^2 \  \delta^3(\sum_{i=1}^3 \textbf{k}_k)~,
	\end{align}
	where the fiducial power spectrum is $\tilde{P}_{\zeta} \equiv 2.1 \times 10^{-9}$. So the clock signal should be proportional to
	\begin{align}\label{cs}
	S^{\mathrm{clock}} \propto f(p,\xi,k_3)\left(\frac{2k_1}{k_3}\right)^{\frac{1+p}{2(1-p)}} \mathrm{sin}\left\{ S\  \mathrm{log}\left[\left(\frac{2k_1}{k_3}\right)^{-1}\right] + \phi(p, \xi, k_3)\right\}~.   
	\end{align}
	We obtain the oscillation phase
	\begin{align}\label{osci}
	S_{\mathrm{oscillation}}  = \mathrm{sin}  \left\{  S \  \mathrm{log}\left[ \left(\frac{2k_1}{k_3}\right)^{-1}\right] + \phi(p, \xi, k_3)\right\}~,
	\end{align}
	which is the essential pattern for the clock signal. We find that this oscillation pattern is the same as the inflation case and we will discuss this point later in detail at the end of Section~\ref{sec3}. It is also important to note the amplitude of the signal as a function of $k_1$, when $k_3$ is fixed. From (\ref{cs}) , the $(k_1/k_3)^{\frac{1+p}{2(1-p)}}$ terms represent how the amplitude changes with the amount of squeezeness, and the function $f(k_3)$ manifests the overall scale  dependence. Generally, $f(k_3)$ can not be determined before we specify a model. 
	From the perspective of $k_3$, these two dependencies are entangled here and we are unable to distinguish them. However, the power of momentum ratio $k_1/k_3$ is determined and can be used to differentiate these scenarios. In practice, we can keep $k_3$ unchanged and increase $k_1$ to study the squeeze shape dependence.

	\subsection{The time-ordered integral}
	The outer and inner integrands of the time-ordered integral take similar forms. We rewrite (2.6) and (2.7) as
	\begin{align}
	\propto \left[\int_{-\infty}^{0} d\tau_1 c_2' a^3 v_{k_3} u_{k_3}^{'*} \int_{-\infty}^{\tau_1} d\tau_2 c_3' a^2 v_{k_3}^{*} u_{k_1}^{'*} u_{k_2}^{'*} \right.
	\end{align}
	\begin{align} \left.+\int_{-\infty}^{0} d\tau_1 c_3' a^2 v_{k_3} u_{k_1}^{'*} u_{k_2}^{'*} \int_{-\infty}^{\tau_1} d\tau_2 c_2' a^3 v_{k_3}^{*} u_{k_3}^{'*} \right]
	\end{align}
	$$+ \mathrm{c.c.} + 2 \ \mathrm{perms}~.$$
	
	To use the saddle point approximation, it is necessary to  make sure that both of the extreme value time $\tau_0$ are in the interval of ($-\infty, \ \tau_1$). In this case, the time-ordered integrals will share the similar oscillating pattern like (3.29). It describes the property of this clock signal, where we can identify the type of alternative scenarios.
	\\
	For simplicity, we consider $\tau_0 = -1$. The modes are denoted as
	\begin{align}
	u_k \propto \frac{1}{(-\tau)^{\frac{p}{1-p}}}  e^{- i k \tau}~,
	\end{align}
	\begin{align}
	v_k \propto \frac{1}{(-\tau)^{\frac{3p-1}{2-2p}}} \left[ C_1 e^{i S \mathrm{log}(-\tau)} + C_2 e^{- i S \mathrm{log}(-\tau)}\right]~. 
	\end{align}
	We denote the variable $x_1 = -\tau_1 $ and $x_2 = - \tau_2$ and the integral (3.30) and (3.31) should be
	$$
	\propto k_1 \int_{0}^{\infty} d x_1 x_1^{\frac{1-3p}{2(1-p)}} \left[C_1^{*} e^{- i S \mathrm{ log}(x_1)} + C_2^{*} e^{ i S \mathrm{log}(x_1)}\right] e^{-2 i k_1 x_1}$$
	
	\begin{align}
	\times \int_{0}^{x_1} dx_2 x_2^{\frac{1+p}{2(1-p)}} \left[C_1 e^{ i S \mathrm{ log}(x_2)} + C_2 e^{ -i S \mathrm{log}(x_2)}\right] e^{- i k_3 x_2}    
	\end{align}
	
	$$
	+ k_1 \int_{0}^{\infty} d x_1 x_1^{\frac{1-3p}{2(1-p)}} \left[C_1 e^{ i S \mathrm{ log}(x_1)} + C_2 e^{ - i S \mathrm{log}(x_1)}\right] e^{-2 i k_1 x_1}$$

	\begin{align}
	\times \int_{x_1}^{\infty} dx_2 x_2^{\frac{1+p}{2(1-p)}} \left[C_1^{*} e^{ - i S \mathrm{log}(x_2)} + C_2^{*} e^{ i S \mathrm{log}(x_2)}\right] e^{- i k_3 x_2}~.  
	\end{align}
	Following the condition of the squeezed limit, we consider $k_3 = 0$ approximately in (3.34) and (3.35). Firstly we deal with the integral (3.34). Compared with the constant mass case, this integral over $x_2$ can be done directly. The second integral term is
	$$
	\left. x_2^{\frac{3-p}{2(1-p)}}\left[C_1\frac{1}{i S + \frac{3-p}{2(1-p)} } x_2^{i S}  + C_2\frac{1}{- i S + \frac{3-p}{2(1-p)} } x_2^{- i S}\right]\right|_{0}^{x_1}   
	$$
	\begin{align}
	\rightarrow x_1^{\frac{3-p}{2(1-p)}}\left[C_1\frac{1}{i S + \frac{3-p}{2(1-p)} } x_1^{i S}  + C_2\frac{1}{- i S + \frac{3-p}{2(1-p)} } x_1^{- i S}\right]~.       
	\end{align}
	There is a similar calculation for the second integral in (3.35). 
	\begin{align}
	\left. x_2^{\frac{3-p}{2(1-p)}}\left[C_1^{*}\frac{1}{-i S + \frac{3-p}{2(1-p)} } x_2^{-i S}  + C_2^{*}\frac{1}{i S + \frac{3-p}{2(1-p)} } x_2^{ i S}\right]\right|_{x_1}^{\propto}~,	
	\end{align}
	which also contains a similar  $x_1$-dependent term like (3.36).
	\begin{align}
	\rightarrow - x_1^{\frac{3-p}{2(1-p)}}\left[C_1^{*}\frac{1}{ - i S + \frac{3-p}{2(1-p)} } x_1^{-i S}  + C_2^{*}\frac{1}{ i S + \frac{3-p}{2(1-p)} } x_1^{ i S}\right]  + ... ~,
	\end{align}
	where the $\langle ... \rangle$ refers to the $x_1$-independent term from the upper infinity case. After combining with the first integral of each term, for (3.34)
	$$
	\rightarrow k_1\int_{0}^{\infty} dx_1 x_1^2 \left[C_1 \frac{1}{iS + \frac{3-p}{2(1-p)}} x_1^{i S} +
	\right.$$
	\begin{align}
	\left. C_2 \frac{1}{-iS + \frac{3-p}{2(1-p)}} x_1^{-i S}\right] \left[C_1^{*} x_1^{-i S} + C_2^{*} x_1^{i S}\right] e^{-2i k_1 x_1}~.
	\end{align}
	
	Only the terms with resonance remains for the contribution to the clock signal, which is proportional to
	\begin{align}
	\rightarrow \int_{0}^{\infty} dx_1 x_1^2 C_1 C_2^{*} \left(\frac{1}{iS + \frac{3-p}{2(1-p)}}\right) \mathrm{exp}\left[  2 i S\  \mathrm{log}(x_1) -  2 i k_1 x_1 \right]~.  
	\end{align}

	In the same way, from (3.35)
	$$
	\rightarrow   - \int_{0}^{\infty} dx_1 x_1^2 C_1 C_2^{*} \left(\frac{1}{iS + \frac{3-p}{2(1-p)}}\right)\mathrm{exp}\left[  2 i S\  \mathrm{log}(x_1) -  2 i k_1 x_1 \right]   $$
	\begin{align}
	+ G(p , \xi) \left\{ k_1 \int_{0}^{\infty} d x_1 x_1^{\frac{1-3p}{2(1-p)}} \left[C_1 e^{ i S \mathrm{ log}(x_1)} + C_2 e^{ - i S \mathrm{log}(x_1)}\right] e^{-2 i k_1 x_1} \right\}~.  
	\end{align}
	We can figure out that the oscillatory terms and coefficients of (3.39) and the first term of (3.40) are consistent with each other. So they can offset each other and only the second of (3.40) remains, which shares the same form with the non-time-ordered integral. To summarize, we can express the time-ordered integral by saddle point approximation as
	\begin{align}
	S^{\mathrm{clock}} \propto (2k_1)^{\frac{1+p}{2(1-p)}} \mathrm{sin}\left\{S \ \mathrm{log}\left[(2k_1)^{-1}\right] + \mathrm{phase} \right\}~.    
	\end{align}

	\begin{figure}[hptb]
		\centering 
		\includegraphics[width=0.8\textwidth]{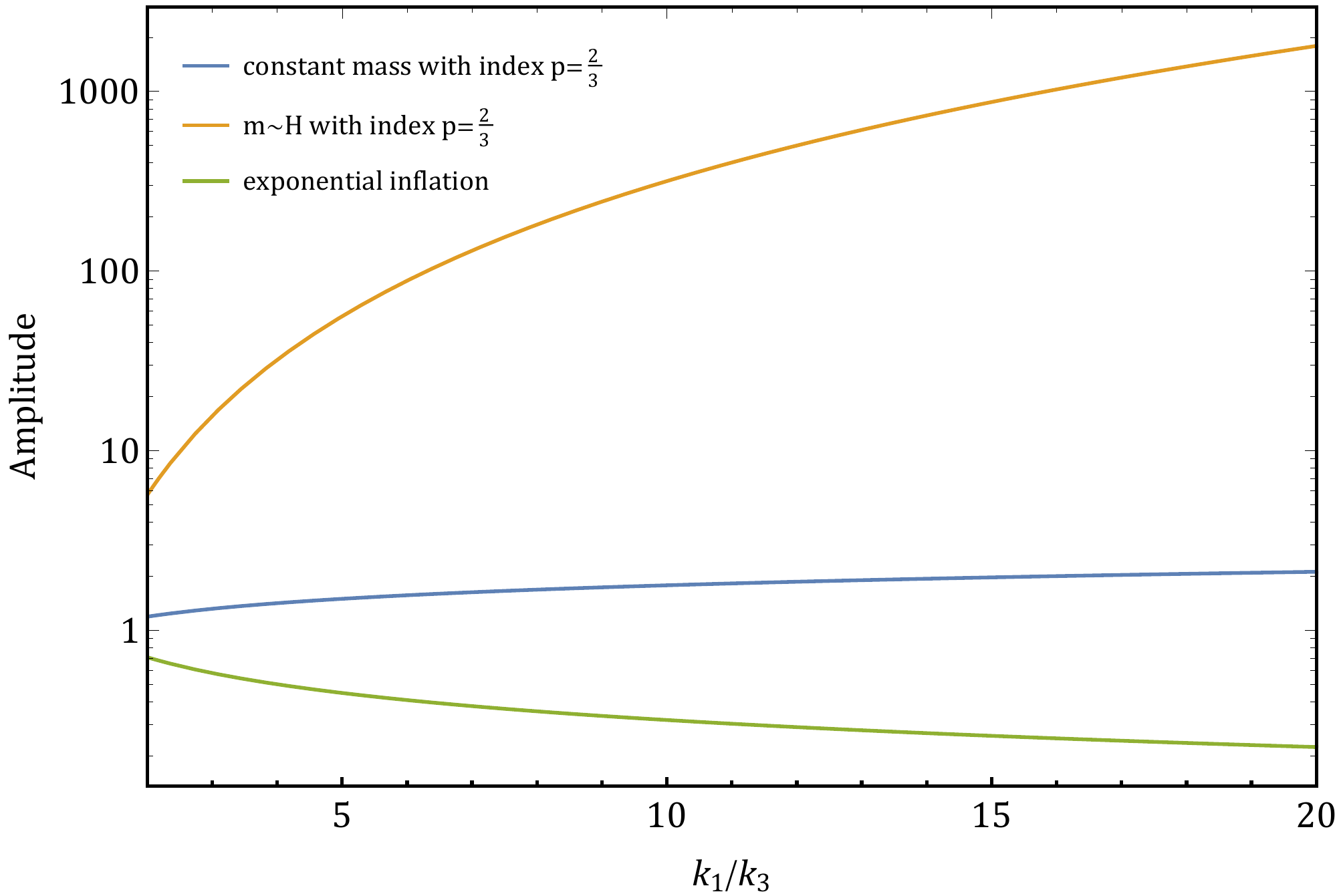}
		\caption{Squeeze shape dependence of amplitude in three different scenarios:  fast contraction ($p$=2/3) with constant mass , fast contraction ($p$=2/3) with mass $\sim H$ and exponential inflation. \label{fig2} }
	\end{figure}

	 Interestingly, this oscillation pattern is the same as the standard clock signal of inflation with constant mass.  There are two intuitive ways to understand this similarity. First of all, the clock signal is generated when the massive field oscillates as $e^{i \int m dt}$ which transformed into conformal time is $e^ {i \frac{m}{H} log(-\tau)}$ in the inflation case. This logarithmic function dependence cause the unique clock signal of inflation. However, the mass generated from Ricci scalar has the property~$m\propto t^{-1}$, this would also introduce  a logarithmic function on the exponent. Another way to understand it is from equation of motion (\ref{masseom}). The mass term combines with the scale factor as $(a m)^2$, in the inflation case, $a m \sim \tau^{-1}$ and in our case $ a m \sim a H \sim \tau^{-1}$. So that, this two different cases share similar equations of motion. As a result, the oscillation pattern of quantum clock signal becomes similar to the standard clocks for inflation.\\

	 Thus, if one uses shape information only to distinguish inflation and alternative scenarios, there is a possibility that alternative to inflation scenarios with dynamically generated mass would be confused with inflation. To break this degeneracy, we note that, although the oscillation patterns are quite similar, the amplitude of different shape functions differs greatly. As we can see from~(\ref{cs}), in addition to the oscillation part, the amplitude of this signal also shows a momentum ratio $k_1/k_3$ dependence. This $ \left({2k_1}/{k_3}\right)^{\frac{1+p}{2(1-p)}}$ term represents how the amplitude changes, when $k_3$ is fixed and one varies $k_1$. For inflation, the corresponding factor is $\sim (k_1/k_3)^{-\frac{1}{2}}$, corresponding to the $p\rightarrow \infty$ limit which differs
	 \footnote{For comparison, for alternative to inflation with constant mass, the corresponding factor is $\sim (k_1/k_3)^{-\frac{1}{2}+\frac{1}{2p}} $, which is also different.}
	 from the case of a finite $p$. We illustrate this difference in Figure~\ref{fig2}. In this non-standard case, one notes that for inflation (or super-inflation where the universe is quasi-de Sitter but with increasing energy density), $|p|>1$ and the oscillation is decreasing while $k_1$ increases. But for alternatives, $-1<p<1$ and the signal is increasing while $k_1$ increases\footnote{In standard case, for slow expansion scenarios ($-1<p<0$), the signal is increasing while $k_1$ decreases.}.

	\section{Time-dependent effective mass from nonlinear corrections} \label{sec4}
	\subsection{The effective mass}
	An effective mass of $\sigma$ may also be generated from interactions in alternative scenarios. We consider the simplest case : $V = \frac{1}{4} \lambda \sigma^4$.
	\\
	Under comparison, it can be separated as $V = \frac{1}{2} \lambda  (3 \langle\sigma^2\rangle) \sigma^2$. We can figure out that the effective mass term should be proportional to $\langle\sigma^2\rangle$, which is related to the two-point correlation function of this scalar field,
	\begin{align}
	\mu^2 = m_{\mathrm{eff}}^2 = 3 \lambda \langle\sigma^2\rangle~,
	\end{align}
	where we can estimate the time-dependent contribution of this two-point correlation in two methods \cite{Linde:2005yw}. The first method is to figure out the expression of quantum fluctuation in alternative scenarios, and do the integration in one cut-off process. In the second method, we separate the scalar in long and short wavelength part and solve them form an nonlinear stochastic different equation (SDE).
	
	The general EOM of scalar field $\sigma$ is
	\begin{align}
	\ddot{\sigma} + 3 H \dot{\sigma} + \frac{k^2}{a^2}\sigma + m^2 \sigma = 0~.
	\end{align}
	The last term should vanishes in this massless case. We combine the scale factor and field as $v = a \sigma$ for simplification. Shown under the conformal time $\tau$, the EOM is
	\begin{align}
	v_k'' + \left(k^2 - \frac{a''}{a} \right) v_k = 0~,
	\end{align}
	which is knowns as the Mukhanov-Sasaki (MS) equation. Here, $v_k''$ and $a''$ denote the second derivative with respect to the conformal time $\tau$.
	\\
	The scale factor $a$ is still expressed as power-law function, under these relations above,
	\begin{align}
	a =a_0 \left(\frac{\tau}{\tau_0}\right)^{\frac{p}{1-p}} \ \ \Rightarrow \ \ \frac{a''}{a} = \frac{(\frac{p}{1-p} - 1)\frac{p}{1-p}}{\tau^2}~.
	\end{align}
	So, the MS equation is
	\begin{align}
	v_k'' + \left(k^2 - \frac{\nu^2 -\frac{1}{4}}{\tau^2} \right) v_k = 0 ~,
	\end{align}
	where $\nu(p) = \frac{3}{2} + \frac{1}{p-1}$ is a model-dependent index.
	The general solution of this MS equation is a linear combination of Hankel functions,
	\begin{align}
	v_k = D_1 \sqrt{-k\tau}H_{|\nu|}^{(1)}(-k \tau) + D_2 \sqrt{-k\tau}H_{|\nu|}^{(2)}(-k \tau)~.
	\end{align}
	This solution should be normalised on the sub-Hubble scales  to constraint constants $D_1$and $D_2$. Under the limit: $- k\tau \rightarrow 0$, the quantum fluctuation $\sigma_k$ is
	\cite{Miranda:2019ara, Nakao:1988yi}
	\begin{align}
	\sigma_k = i \frac{1}{\sqrt{4 \pi k a^2}}\frac{2^{|\nu|} \Gamma(|\nu|)}{(-k\tau)^{|\nu|-\frac{1}{2}}}~.
	\end{align}
	The time-dependent contribution of quantity $\langle\sigma^2 \rangle$ is proportional to
	\begin{align}
	\int d^3 k \ \frac{1}{a^2} \ k^{-2|\nu|} (-\tau)^{1- 2|\nu|}~.
	\end{align}
	We use the time dependent physical momentum $p = \frac{k}{a}$ to emphasize the physical meaning,
	\begin{align}
	\langle \sigma^2 \rangle \propto \int_0^H dp\  p^{2-2|\nu|} \ (-\tau)^{1-2|\nu|} \ a^{1-2|\nu|}~.
	\end{align}
	Since $(-a \tau) \propto H^{-1}$ and the time dependent Hubble constant $H = \frac{\dot{a}}{a} = \frac{p}{t}$ \cite{Linde:2005ht},
	\begin{align}
	\langle \sigma^2 \rangle \propto H^{3-2|\nu|} H^{2|\nu|-1} = H^2 \propto \frac{1}{t^2}~.
	\end{align}

	There is a more accurate method to obtain the time-dependent contribution. We still consider the alternative models with the scalar field $\sigma(x, t)$. 
	The idea is to separate this scalar field using the stochastic approach. We split the $\sigma$ field into the long wavelength part $\sigma_L$ and short wavelength part $\sigma_S$ with the expression
	\begin{align}
	\sigma(x, t) = \sigma_L + \sigma_S = \sigma_L + \int \frac{d^3 k }{(2 \pi)^{3/2}} \Theta[k - \epsilon a(t) H]\left[a_k \sigma_{S, k}(t) e^{-i k x} + a_k^{\dagger} \sigma_{S, k}^{*} e^{i k x}\right]~,
	\end{align}
	where $\epsilon$ is a small constant and $a_k$ and $a_k^{\dagger} $ are annihilation and creation operator. $\Theta(x)$ is the Heaviside step function. The short wavelength part satisfies the massless EOM of scalar equation (4.2) with $m^2 = 0$ so $\sigma_{S, k} = \sigma_k$ in (4.7).\\
	The long wavelength part $\sigma_L$ satisfies the classical equation of motion $\ddot{\sigma_L} + 3H \dot{\sigma_L} + \partial_i V = 0$. For the alternative scenarios with index $p$,
	\begin{align}
	\ddot \sigma_L + 3 \frac{p}{t} \dot \sigma_L + \lambda \sigma_L^3 = 0~.
	\end{align}
	We are interested in the damping dominated regime. In this special period, the damping term is much larger than the accelerating term and then we can ignore the second derivative term $\ddot{\sigma_L}$, i.e. 
	\begin{align}
	\ddot \sigma_L \ll 3 H(t) \dot\sigma_L = 3 \frac{p}{t} \dot \sigma_L~,
	\end{align}
	which requires
	\begin{align}
	\lambda t^2 \ll (1-p)\frac{\sqrt{3p}}{\sigma_L}~.
	\end{align}
	Using the Starobinsky's stochastic method, the short wavelength contribution can be added to the original classical equation as a perturbation term $f(x, t)$. The quantum-corrected equation of motion is \cite{Liu:2015dda}
	\begin{align}
	3 \frac{p}{t} \dot{\sigma} = -\lambda \sigma^3 +3 \frac{p}{t} f(x, t)~.
	\end{align}
	The stochastic random force term $f(x, t)$ is
	\begin{align}
	f(x, t) = (p-1) D_1 t^{p-2} \int \frac{d^3 k}{(2\pi)^{3/2}} \delta(k - D_1 t^{p-1})\left[a_k \sigma_{S, k}(t) e^{-i k x} + a_k^{\dagger} \sigma_{S, k}^{*} e^{i k x}\right]~,
	\end{align}
	where $D_1 = \epsilon a_0 t_0^{-p} p$ is a combined constant. Here, the integration is constrained only for specific $k$ because of the delta function. $f(x, t)$ is a Gaussian random field and the two-point correlation function is
	\begin{align}
	\langle f(x_1, t_2), f(x_2, t_2) \rangle = \Lambda t^{-3} \delta(t_1 - t_2)\  j_0\left(\epsilon a(t) H|x_1 - x_2|\right)~,
	\end{align}
	where $j_0(x) = \frac{sin(x)}{x}$ and $
	\Lambda$ is a time-independent combined coefficient \cite{Starobinsky:1994bd}.
	\\
	The quantum-corrected equation of motion then becomes one nonlinear stochastic differential equation (SDE) 
	\begin{align}
	\dot{\sigma} = - \frac{\lambda}{3p} t \sigma^3 + \frac{\sqrt{\Lambda}}{3p} t^{-\frac{3}{2}} \eta(x, t)~,
	\end{align}
	where $\eta(x, t)$ is the simplified function with $\langle \eta(x, t) \rangle = 0 $ and $\langle\eta(x_1, t_1),  \\ \eta(x_2, t_2)\rangle= \delta(t_1 - t_2) \delta(|x_1 - x_2|)$. Because of the time and field-dependent drift coefficient $-\frac{\lambda}{3p} t \sigma^3 $ and time-dependent diffusion coefficient $\frac{\Lambda}{18 p^2} t^{-3}$, we cannot figure out the analytic solution of this nonlinear SDE. The numerical solution shows that the proportion to $t^{-2}$ is still an appropriate approximation for the two-correlation function $\langle \sigma^2 \rangle$. \\
	In conclusion, the effective mass in this case is expressed as
	\begin{align}
	\mu^2 = m_{\mathrm{eff}}^2 \propto \frac{1}{t^2}~.
	\end{align}
	And the effective mass and the potential term become
	\begin{align}
	\mu^2 = 2 \frac{C(p)}{t^2},\ \ 
	V = \frac{C(p)}{t^2} \sigma^2~,
	\end{align}
	where $C(p)$ is a time-independent coefficient which is model-dependent. Approximately, by the first method, 
	\begin{align}
	C(p) = 3\lambda \frac{\Gamma^2(|\nu|) \ 2^{2|\nu|}}{64 \pi^4}~,
	\end{align}
	for the interacting coefficient $\lambda > 0$, $C(p)$ and the potential are always non-negative, which is different from the Ricci-dominated case. Note that there is an undefined scenario for $p = \frac{1}{3}$ ($\nu = 0$). This is the only case leading to an infinite Gamma function so this contracting scenario cannot be dealt with in this way.

	\subsection{The clock signal}
	Since the effective mass is proportional to $t^{-2}$, the clock signal dominated by the quantum fluctuation should be similar to the Ricci-dominated case. Like (3.6), the conformal time basis EOM of scalar field mode $v_k$ is
	\begin{align}
	\frac{d^2 v_k}{d \tau^2} + \frac{2p}{(1-p)\tau}\frac{d v_k}{d \tau} + \frac{2 C(p)}{(1-p)^2 \tau^2} = 0~.
	\end{align}
	The general solution is
	\begin{align}
	v_k = (-\tau)^{\frac{1-3p}{2-2p}}\left[e^{i B(p) log(-\tau)}C_1 + e^{-i B(p) log(-\tau)}C_2\right]~,
	\end{align}
	\begin{align}
	B(p) =  \frac{\sqrt{8 C(p) - (1- 3p)^2}}{2(p-1)}~,
	\end{align}
	which has the same form as (3.15). There is also an restriction on $p$, like the discussion in the last section since $B(p)$ must be real. In the same method like (3.28), we can figure out the clock signal in this case should be proportional to
	\begin{align}
	S^{\mathrm{clock}} \propto (2k_1)^{\frac{1+p}{2(1-p)}} \mathrm{sin}\left\{ B(p)\ \mathrm{log}\left[ \left(2 k_1\right)^{-1}\right] + \mathrm{phase} \right\}~,
	\end{align}
	where C($p$) follows (4.20) expression with $\nu = \frac{3}{2}+ \frac{1}{p-1}$.
	
	\section{Conclusion} \label{con}
	In this paper, we extend the study of quantum primordial non-standard clocks to alternative scenarios with dynamically generated mass. Such time-dependent mass can naturally be generated where the background significantly break scale invariance. Two mechanisms are considered:
	
	The first mechanism is through non-minimal coupling with Ricci scalar $R$ which should be a constant ($\propto H^2$) in de-sitter space-time, but it can contribute a time-dependent effective mass in different alternatives to inflation scenarios with $t^{-2}$ time dependence. \\
	
	We note that not all alternatives to inflation scenarios can produce non-trivial quantum primordial non-standard clocks signal from coupling to $R$. For generating oscillatory signals, the mode function of massive fields should have oscillation behaviors, this condition places a constrain on the power-law index $p$ as a function of $\xi$, as plot in Figure~\ref{fig1}. \\
	
	The second mechanism of dynamical mass generation is from quantum loop correction, we give a simple approximation of this method's contribution. For the  massless field with $\lambda \sigma^4$ interaction, we show that the time dependence of effective mass is also proportional to $ H^2$. \\
	
	Finally, the oscillation behavior of this situation is quite similar to inflation scenarios with the time-independent mass case, because massive fields in these two different scenarios share the similar equation of motion. However, we can still distinguish them from the shape dependence of the amplitude. The amplitude of shape function in our case as  $(k_1/k_3)^{\frac{1+p}{2(1-p)}}$ when $k_1$ varies, which is a decreasing function for inflation and an increasing function for alternative scenarios. It is interesting to see if similar observations can be used to distinguish classical standard and non-standard primordial clocks.

	\section*{Acknowledgement}
	We thank Xingang Chen and Siyi Zhou for fruitful discussions and comments. This work was supported in part by ECS Grant 26300316 and GRF Grant 16301917 and 16304418 from the Research Grants Council of Hong Kong.

	\appendix
	\section{General expression of the clock field}\label{eom}
	In the above discussion, we consider the small $m$ limit, where the effective mass is dominated by the Ricci curvature. Here we show a general case for the equation of motion of the massive scalar field under the conformal time,
	\[
	\frac{d^2 v_k}{d \tau^2} + \frac{2p}{(1-p) \tau} \frac{d v_k}{d \tau} + [k^2 + 6\xi\frac{p(2p-1)}{(1-p)^2 \tau^2} + m^2 \frac{a_0^2}{t_0^{2p}} \tau^{\frac{2p}{1-p}}]v_k = 0~. 	
	\tag{A.1}
	\]
	For the massless or relativistic case with $k \gg ma$, the EOM is
	\[
	\frac{d^2 v_k}{d \tau^2} + \frac{2p}{(1-p) \tau} \frac{d v_k}{d \tau} + [k^2 + 6\xi\frac{p(2p-1)}{(1-p)^2 \tau^2} ]v_k = 0
	\tag{A.2}~.
	\]
	The analytical solution is:
	\[
	v(\tau) \rightarrow (2p-2)^{\frac{1-3p}{2-2p}} (-\tau)^{\frac{1-3p}{2-2p}}[C_1 \mathrm{BesselJ}(P(p,\xi),  k\tau) +C_2 \mathrm{BesselY}(P(p, \xi), k\tau) ]~,
	\tag{A.3}
	\]
	where 
	\[
	P(p, \xi) = \frac{\sqrt{1-6p + 9p^2 + 24p \xi - 48p^2 \xi}}{2p-2}~.
	\tag{A.4}
	\]

	\section{Saddle point approximation}\label{apr}
	We use the saddle point approximation to deal with the complicated integrals in (2.5) and (2.6). If the integrated term contains fast-oscillating part $e^{i g(t)}$ and weak time-dependent function $f(t)$, the integral can be approximated by the saddle point approximation as
	\[
	\int_{a}^{b} dt\  e^{i g(t)} f(t) \approx e^{i g(t_0) \pm i \frac{\pi}{4}} f(t_0) \sqrt{\frac{\pm 2 \pi}{g''(t_0)}}~. \tag{A.5}
	\]
	Here, $t_0$ is the time leading to the extreme value of oscillating function $g(t)$, i.e. $g'(t_0) = 0$. The choices of plus or minus signs are for computing convenience which both lead to the same results.
	\section{Single variable Langevin Equation}
	The Langevin equation with a stochastic variable $\phi$ takes the form
	\[
	\dot{\phi} = f(\phi, t) + g(\phi, t) \Gamma(x,t)~,
	\tag{A.6}
	\]
	where the Langevin force term $\Gamma(t)$ satisfies Gaussian random distribution with
	\[
	\langle \Gamma(x, t) \rangle=0,\ \ \langle \Gamma(x, t)\Gamma(x', t') \rangle = \delta{(x-x)}\delta{(t-t')}~.
	\tag{A.7}
	\]
	For nonlinear Langevin equations, generically we cannot obtain analytic solutions. However, for special cases like $g(\phi, t) = g(\phi)$, there could be tricks to transfer the form to a solvable one. We can also solve this by the Fokker-Planck equation to get the probability density $W(\phi, t)$ \cite{Risken}. After that, the correlation function is
	\[
	\langle h(\phi) \rangle = \int_{-\infty}^{\infty} h(\phi) W(\phi, t) d\phi~.
	\tag{A.8}
	\]

\end{document}